\documentclass[12pt]{article}

\usepackage[margin=1in]{geometry}
\usepackage{hyperref}
\usepackage[utf8]{inputenc}
\usepackage{csquotes}
\usepackage{mathtools}
\usepackage{amssymb}
\usepackage{amsmath}
\usepackage{upgreek}
\usepackage{epigraph}

\usepackage[backend=biber]{biblatex}
\usepackage{graphicx}
\graphicspath{ {./images/} }
\usepackage{caption}

\newcommand{\La}{\mathcal{L}}

\newcommand{\Ta}{\mathcal{T}}

\newcommand{\Ga}{\mathfrak{S}}

\newcommand{\?}{\stackrel{?}{=}}
\newcommand{\Marr}{\underset{M}{\Rightarrow}}
\newcommand{\SIGMA}{\boldsymbol{\upsigma}}

\author{Jonathan J. Mize \footnote{Philosophy \& Mathematics, University of North Texas. Email: \href{mailto:jonathanmize@my.unt.edu}{jonathanmize@my.unt.edu}}}
\title{Machines as Programs\\ P $\neq$ NP}
\date{}

\begin{document}

\maketitle

\begin{abstract}

The Curry–Howard correspondence is often called the proofs-as-programs result. 
I offer a generalization of this result, something which may be called machines as programs. 
Utilizing this insight, I introduce two new Turing Machines called ``Ceiling Machines." 
The formal ingredients of these two machines are nearly identical. 
But there are crucial differences, splitting the two into a ``Higher Ceiling Machine" and a ``Lower Ceiling Machine." 
A potential graph of state transitions of the Higher Ceiling Machine is then offered. 
This graph is termed the ``canonically nondeterministic solution" or \textit{CNDS}, whose accompanying problem is its own replication, i.e., the problem, ``Replicate \textit{CNDS}" (whose accompanying algorithm is cast in Martin–L\"of type theory). 
I then show that while this graph can be replicated (solved) in polynomial time by a nondeterministic machine—of which the Higher Ceiling Machine is a canonical example—it cannot be solved in polynomial time by a deterministic machine, of which the Lower Ceiling Machine is also canonical. 
It is consequently proven that P $\neq$ NP. 

\end{abstract}

\section{Preamble to Proof}

Professor of computer science Oded Goldreich has claimed that he will ``refuse to check claims regarding the resolution of famous open problems such as P versus NP [...] unless the claim is augmented by a \emph{clear and convincing indication} as to how this work succeeds where many others have failed.”
As homage to Dr. Goldreich—and numerous others who, no doubt, share his sentiment—I will first say a few words on \emph{how this work succeeds where many others have failed.}

According to an unofficial source, there have been 116 total “proofs” of P versus NP.\footnote{\href{https://www.win.tue.nl/~gwoegi/P-versus-NP.htm?fbclid=IwAR0Glw5o2U6h4HCTizR7LoXCeaiiIPf19W0_BPBLwHhTsDzaefFrfdyksi4}{``The P-versus-NP page''} (GJ Woeginger)}
Whatever the precise number may be, one thing is for certain: a great deal of theoretical “rubber” has been burnt, with little explanatory traction to show for it. 
Now, I obviously cannot go down the line of proposed proofs and pick out each one’s ``fatal flaw" (perhaps to do so would require a nondeterministic Turing Machine!). 
However, I can show how my proof differs from the others \emph{dramatically}.

Although I can say with no certainty that my approach is the only feasible one, I can emphasize that it \emph{is} the \emph{most general possible solution} to the problem. 
Obviously, this is a tremendous claim.
And it must be backed up. 
First, I have to precisely formulate what exactly I mean by ``general." 
Certainly, I will not be able to provide a formal proof to the effect that this much is the case, but I can impart the basic intuition behind the assertion.

Consider Tarski’s ``hierarchy” of \emph{metalanguages} and their corresponding \emph{object languages}. 
Recall that the definitions in a given object language $L$ are necessarily given in a metalanguage of $L$, $M$. 
Anything that can be said in $L$ can be said in $M$, but not the converse.
From these simple notions (among others), Tarski was able to define truth in a formal system ``from without," or by invoking semantic notions such as ``meaning"" or “denoting” \emph{only} within the metalanguage $M$ and not the object language $L$. 
Consider the following:
\begin{displayquote}
	For all $x$, $True(x)$ if and only if $\phi$(x), \\
	where the predicate $True$ does not occur in $\phi$.
\end{displayquote}
A hierarchy of languages can then be seen as a sort of transitive ordering \{$M_{1} > M_{2} > ... > M_{n} > L$\} in which the relation $>$ expresses that the first language can ``talk about" the second language (and all subsequent languages) but not the converse.

This \emph{metamathematical} notion perfectly captures the intuition behind supposed ``generality” of solution. 
Na\"ively , we can say that a given solution to a problem $S_{i}$ is ``more general than” another $S_{j}$ , just in case that $S_{i}$ is formulated in a language $L_{i}$ which can ``talk about” the language in which $S_{j}$ is formulated in, $L_{j}$. 
So, when I claim that my solution to the problem is ``more general” than other proposed solutions, what I mean is that I utilize just such a language $L_{i}$ which can ``talk about” the language(s) in which the other solutions are formulated. 
Of course, now, we must be precise about which ``languages” we are talking about exactly.

Superficially, a majority of the supposed 116 ``proofs” of the problem are formulated mathematically. 
This is not exactly what I mean when I say ``language.” 
Quite obviously, a great many mathematical techniques are required to solve such a problem (and I make use of many myself). 
What I have in mind with ``language” can perhaps better be called ``governing paradigm.” 
Forgive me if that sounds too high-flyingly philosophical. 
Although the term cuts straight to the issue, perhaps something like ``canonical object language” would be more appropriate.

Although buzzwords like \emph{governing paradigm} and \emph{canonical object language} might sound forbidding or needlessly wordy, the basic idea behind it all is rather simple. 
Consider that in any branch of mathematics there are \emph{established ways of doing things}, \emph{canonical methods} and structures and so forth. 
Computational complexity theory is no different, neither is the theory of computation in general.

The new ``paradigm” I offer I term \emph{machines as programs}. 
This approach is inspired by the Curry–Howard correspondence, often termed the ``proofs-as-programs” result.

Where a traditionally conceived Turing Machine consists of states, a state transition function, an input alphabet, etc., a machine adhering to the machines-as-programs maxim simply adjoins to such a machine a programmatic extension $\langle\Ta, A, \La\rangle$. As far as $\langle\Ta, A, \La\rangle$ is concerned, $\Ta$ holds theorems, $A$ holds states and input symbols and $\La$ holds all the requisite symbols for $\Ta$. 
Superficially, this 3-tuple appears as a logical structure, \emph{sans} interpretation function. 
But this ``programmatic extension” cannot be reduced to such a thing (neither can it be reduced to mere model-theoretic construction). 
Why is this so? 
For this simple reason: $\Ta$ and $\La$ are not merely abstract mathematical objects, but they are \emph{stored data} in a $CM$. 
More specifically, $\Ta$ and $\La$ are ``activated” by input into $A$. 
In this manner, programs of a $CM$ are instances of $A$-transitions which conform to the stored data in $\Ta$ and $\La$.

Moreover, we can adjoin any generic $n$-tuple machine to such a programmatic extension, simply by expanding the $\La$ of $\langle\Ta, A, \La\rangle$ such that $\La$’ = $\La \cup M$ for any arbitrary definition of $M$ (say $M$ = \{$Q, \Gamma, b, \Sigma, q_{0}, A, \delta$\}). 
In this manner, we will build up to two different machines, one with only \emph{deterministic} machines (\textit{DTM}) in its expansion ($\La$’ = $\La \cup DTM_{i}, ..., DTM_{n}$) and another with only \emph{nondeterministic} machines in said expansion, ($\La$’ = $\La \cup NTM_{i}, ..., NTM_{n}$). 
The former we will call the Lower Ceiling Machine and the latter the Higher Ceiling Machine.

This is \emph{machines as programs} in action.

This conceptual innovation—allowing machines to accept ``expansions” of other machines—is crucial to establishing the language $L_{i}$, corresponding to the machines-as-programs approach, as the “most general” manner of attack in the Tarskian sense discussed above.. 
This ``innovation” centers on one crucial goal (among many others)—a solution to the dilemma first proposed by Baker, Gill, and Solovay in 1975.

In their paper, ``Relativizations of the P $\?$ NP Question,” these gentlemen discovered something troubling for computational complexity theorists: so-called ordinary diagonalization methods of proof, when taken to apply \emph{equally as well} to two \emph{relative} classes, would imply a non-solution to \emph{all} such ``relativized” P $\?$ NP questions. 
Baker, Gill and Solovay went on to claim that ``[our] results suggest that the study of natural, specific decision problems offers a greater chance of success in showing P $\neq$ NP than constructions of a more general nature.” 
This was a massive blow to hopeful computational complexity theorists everywhere. 
It seemed that—whatever technical tools were utilized to finally prove the problem—a sort of ``hunt and peck” method would need to be utilized. 
Thankfully, I contend, a general solution to the problem is alive and well. 
And whatever Baker, Gill and Solovay had in mind when they mentioned ``constructions of a [...] general nature,” they did not bother to consider the possibility of a general solution capable of \emph{obviating the issues associated with relativization}.

Interestingly enough, a hint to the dilemma lies in Baker, Gill and Solovay’s very own paper. 
In their introduction, the theorists stated the following:
\begin{displayquote}
	By slightly altering the machine model, we can obtain differing answers to the relativized question. 
	This suggests that \emph{resolving the original question requires careful analysis of the computational power of machines}. (emphasis added)
\end{displayquote}
In other words, the particular machine model of any given machine is likely not as important as once thought to the solution of P $\?$ NP. 
And, indeed, my proposed general solution to the problem conforms to this intuition of Baker, Gill and Solovay. 
As it so happens, a higher-order analogue of \emph{formal software verification} factors into my solution in a prominent way.

Where any given formal software verification problem consists of formally establishing that a given \emph{implementation} satisfies a \emph{specification}, an interesting ``angle of attack” is to designate two machines, the so-called Higher Ceiling Machine from earlier (a \emph{nondeterministic} machine) and the so-called Lower Ceiling Machine (a \emph{deterministic} one) and consider the former to be a \emph{specification} and the latter to be an \emph{implementation}. 
Informally:
\begin{displayquote}
	\begin{center} \textit{NTM} $\Rightarrow$ \textit{specification} \\
	\textit{DTM} $\Rightarrow$ \textit{implementation} \end{center}
\end{displayquote}
In essence, the \textit{NTM} acts as a ``template” or an abstracted structural description of an “ideal software.” 
And, if the \textit{DTM} is to solve a problem characteristically solvable in polynomial time by a nondeterministic machine, it must conform to this \emph{ideal software} of \textit{NTM}.

In the ``real world,” however, it is far from obvious what such a thing could mean. 
How can we even talk about the “ideal software” of a nondeterministic machine, if we have no idea what such a thing could look like, in the \emph{real world}? 
This is an understandable qualm, but there are, no doubt, rigorous means of skirting around it. 
We can actually abstract over the issue of what exactly a real-world piece of \textit{NTM} software could “look like,” and simply consider what sort of \emph{logical rules} it would obey. 
Consider this a sort of \emph{formal software structuralism}.

When we view \textit{HCM} and \textit{LCM} in this light, an interesting question to ask is this: how does this \emph{specification-implementation} metaphor “translate” into talk of the respective \emph{state transitions} of \textit{HCM} and \textit{LCM}? 
A deeper question still is this: if \textit{LCM} cannot quite “conform” to the specification that is \textit{HCM} (as could be expected as a nondeterministic machine could possesse fundamentally “different” software than a deterministic one), what ramifications would this have on the ability of LCM to \emph{reproduce} the \emph{state transitions} of HCM as \emph{input query}? 
Although the technical details of this portion of the proof are quite involved, I can offer a good bit of intuition behind the specific maneuvers.

Here is perhaps the most important “trick”—\emph{fashion a solution whose leading algorithm replicated it}. 
But, without context, this likely sounds inscrutably obscure, so let me parse this. 
First, let us consider the following:
\begin{displayquote}
	\begin{center} \{$(a^{\sigma} \times a^{q})^{i} \times (a^{\sigma} \times a^{q})^{j} \times \cdots (a^{\sigma} \times a^{q})^{n}$\} = a polynomial-time computation produced by \textit{HCM}

	\textbf{REP} = an algorithm designed to replicate the above computation

		\{($a^{\sigma} \times a^{q})^{i} < \cdots < (a^{\sigma} \times a^{q})^{n-x}$\}, with accompanying intermediary sets $x$ = a solution produced by \textbf{REP}
	\end{center}
\end{displayquote}
The first item of above, the computation produced by \textit{HCM}, we call the canonically nondeterministic solution or \textit{CNDS}. 
What we can do is associate with \textit{CNDS} the decision problem or “query”, $\textit{CNDS}^{Q}$, outputting an affirmative answer if the computation produced by \textit{HCM} can be reproduced by \textit{LCM} and a negative answer if it cannot be. 
Recall the statement from above, “a machine adhering to the machines-as-programs maxim simply adjoins to such a machine a programmatic extension $\langle\Ta, A, \La\rangle$.” 
Each $\Ta$, of both the \textit{HCM} and \textit{LCM} contains programs as stored in memory (or program templates).

The gist of our proof is quite simple. 
We simply prove that the \textit{LCM} does not contain the requisite members of $\Ta$ to carry out the algorithm \textbf{REP}, in order to answer the query $\textit{CNDS}^{Q}$. 
And, as an \textit{LCM} is proven equivalent to a deterministic machine and \textit{HCM} equivalent to a nondeterministic one, we thus have an example of a query, $\textit{CNDS}^{Q}$ , which \emph{can} be trivially answered by a \emph{nondeterministic} machine but \emph{cannot} be answered by a \emph{deterministic} one (an \textit{LCM} in this case). 
It is consequently proven that P $\neq$ NP.

\section{Building a Ceiling Machine}
\textbf{Definition 1.} The main formal device we will require is \emph{intuitionistic type theory}, also known as \emph{Martin–L\"of type theory}. 
We will make use of the following canonical interpretation of logical constants as type formers:
\begin{align*} 		\text{$\bot$} &=  \text{$\varnothing$} \\
			\text{$\top$} &=  \text{$1$} \\
			\text{$A \lor B$} &=  \text{$A + B$} \\
			\text{$A \land B$} &=  \text{$A \times B$} \\
			\text{$A \supset B$} &= \text{$A \rightarrow B$} \\
			\text{$\exists$$x$:$A$. $B$} &= \text{$\Sigma$$x$:$A$. $B$} \\
			\text{$\forall$$x$:$A$. $B$} &= \text{$\Pi$$x$:$A$. $B$} \end{align*} 
$\Sigma$$x$:$A.$ $B$ is the \emph{disjoint sum} of the $A$-indexed family of types $B$, while $\Pi$$x$:$A.$ $B$ is the \emph{cartesian product} of the $A$-indexed family of types $B$. Elements of $\Sigma$$x$:$A.$ $B$ are pairs ($a, b$) and the elements of $\Pi$$x$:$A.$ $B$ are computable functions $f$.

When we prove sentences or theorems in Martin–L\"of type theory we build a \emph{construction} or \emph{proof-object} which “witnesses” the truth of the sentence or theorem in question. 
From here on, I will refer to such a construction simply as a proof-object. \begin{displayquote} \end{displayquote}
\textbf{Remark 1.1.} In what follows, we will make great use of the \textit{Curry–Howard correspondence}. Informally, this result casts \emph{proofs as programs}. 
Although there have been numerous syntheses of this result, all we will require for our result is Martin–L\"of type theory.

However, I do not intend to rely upon only the Curry–Howard result as it is traditionally conceived. 
I would also like to show another possible “rendition” of this correspondence: \emph{machines as programs}. 
No, I do not intend to reduce all machines to mere “coding.” 
Neither do I intend to discount the structural properties of the wonderfully diverse computing machines.

Perhaps the best way to put it is like so: the machines-as-programs result allows us to \emph{embed} any well-defined machine in a \emph{programmatic extension} of said machine. 
By “programmatic extension” I mean that every well-defined function of said “embedded” machine can be executed through the \emph{programming} of another, “higher” machine. 
Just what is this “higher machine”? 
For now, we might as well call it a \textit{Ceiling Machine} or \textit{CM} for short.

Simplistically, a Ceiling Machine is just a 3-tuple addition, $\langle\Ta, A, \La\rangle$ to the usual conception of a Turing Machine. 
As far as $\langle\Ta, A, \La\rangle$ is concerned, $\Ta$ holds theorems, $A$ holds states and input symbols and $\La$ holds all the requisite symbols for $\Ta$. 
Superficially, this 3-tuple appears as a logical structure, \emph{sans} interpretation function. 
But this “programmatic extension” cannot be reduced to such a thing (neither can it be reduced to mere model-theoretic construction). 
Why is this so? 
For this simple reason: $\Ta$ and $\La$ are not merely abstract mathematical objects, but they are \emph{stored data} in a \textit{CM}. 
More specifically, $\Ta$ and $\La$ are “activated” by input into $A$. 
In this manner, programs of a CM are instances of $A$-transitions which conform to the stored data in $\Ta$ and $\La$. 
This is \emph{machine as program} in action.  \begin{displayquote} \end{displayquote}
\textbf{Definition 1.2.} For now, we might as well define a \textit{Ceiling Machine} as the 4-tuple $\langle\Ta, A, \La, \delta\rangle$. 
We said above that \emph{machines as programs} “allows us to \emph{embed} any well-defined machine in a \emph{programmatic extension} of said machine.” 
We can now expand on this. 
Take the 4-tuple $\langle\Ta, A, \La, \delta\rangle$ of any generic \textit{CM} and \emph{expand} $\La$ of \textit{CM} by the symbols of the generic 6-tuple \{$Q, \Gamma, b, \Sigma, q_{0}, A$\} of machine $M$, such that $\La$’ = $\La \cup M$. 
This preserves the intuition that any generic tuple like \{$Q, \Gamma, b, \Sigma, q_{0}, A$\} is really just a language whose definitions inhabit a corresponding theory $\Ta$.

Note that any $M$ of the expansion $\La$’ = $\La \cup M$ of a \textit{CM} is such that \emph{the transition function or relation is omitted}. 
The transition function/relation in the extension $\langle\Ta, A, \La, \delta\rangle$ of a \textit{CM} serves as the governing function/relation for each embedded machine $M$ of \textit{CM}. 
This aspect of \textit{CM}s will be extremely important for our main results.

Denote $\langle\Ta, A,$ $\La$$’, \delta\rangle$ as a \textit{CM} which has $\La$’ = $\La$ $\cup$ $M$ by any arbitrary tuple corresponding to $M$. 
Call $M$ in this case an \emph{embedded machine} of \textit{CM}. 
For our results, we assume that each expansion corresponds to an appropriate expansion of theory, $\Ta$’ = $\Ta \cup M$. 
A theory of an arbitrary $M$ simply constrains the potential \emph{behavior} of $M$ (possible input, state transitions, etc.). 
The specifics behind expansion of theory are not material to the logic of the proof.  \begin{displayquote} \end{displayquote}
\textbf{Definition 1.3.} We can now expand on what exactly it means to be a \emph{program} of a \textit{CM}. 
The Curry–Howard correspondence allows us to define programs as \emph{proofs} in Martin–L\"of type theory. 
We said above that $\Ta$ and $\La$ are \emph{not} merely abstract mathematical objects, but they are \emph{stored data} in a \textit{CM}. 
We also noted that $\Ta$ and $\La$ are “activated” by input into $A$, thus entailing that, programs of a \textit{CM} are instances of $A$-transitions which conform to the stored data in $\Ta$ and $\La$. 
This makes a proof in a \textit{CM} the following: a transition of the schema ($A \rightarrow A \rightarrow \cdots A$) such that this string corresponds to a sequence of \emph{derivable formulae} in $\Ta$ comprising members of $\La$. 
In this manner, each \emph{derivation step} of any given \textit{CM} program must be already stored in $\Ta$. 
Thus, for any given proof-object $p_{i}$ of $\Ta$, $\Ta$ must also contain $j_{1}, ..., j_{i-1}$ of judgements such that $j_{1}, ..., j_{i-1}, p_{i}$. 
Call this property of \textit{CM}’s $\Ta$ \emph{derivational closure}. 
Note that any given $p_{i}$ may correspond to multiple sets of judgements such that $j_{1}, ..., j_{i-1}$, $p_{i}$. 
It is immaterial to our main results exactly which judgement sets any given $p_{i}$ contains. 
More precisely speaking then, a proof/program in any generic \textit{CM} is the following:
\begin{displayquote} \begin{center}
	a string ($a_{i} \rightarrow a_{j} \rightarrow \cdots a_{n}$) of the schema ($A \rightarrow A \rightarrow \cdots A$),

such that sets of ($a_{i}, ..., a_{n-x}$) of ($a_{i} \rightarrow a_{j} \rightarrow \cdots a-{n}$) correspond to combinations of $j$ and $p$ of some sequence of derivations ($j_{1}, ..., j_{i-1}, p_{i}$) where each $j$ and $p$ of ($j_{1}, ..., j_{i-1}, p_{i}$) is a formula $\varphi$ in \textit{CM}’s $\Ta$\footnote{More precisely, this is a \emph{computation}. 
But we will refer to such computations as strings, hopefully without too much confusion.} \end{center}

\end{displayquote}
\textbf{Definition 1.4.} We stated above that $A$ holds states and input symbols. 
But this is rather vague. 
All that was meant by this cryptic-sounding remark is that, per any \textit{CM}, we have \emph{no reason} to distinguish between input symbols and states of the \textit{CM}. 
Accordingly, we can offer the following transition function for a \textit{CM}:
\begin{displayquote} \begin{center}
	$\delta$: ($A^{\sigma} \times A^{q}$) $\rightarrow$ ($A^{\sigma} \times A^{q}$) $\times$ \{L, R\}, \\
where traditionally conceived \emph{input} $\sigma$ is accompanied by traditionally conceived \emph{state} $q$. \end{center}
\end{displayquote}
Thus, where we talk about a string ($a_{i} \rightarrow a_{j} \rightarrow$ $\cdots$ $a_{n}$) of the schema ($A \rightarrow A \rightarrow \cdots A$), it is more precise to talk about a string \{($a^{\sigma} \times a^{q})^{i}$ $\rightarrow$ ($a^{\sigma} \times a^{q})^{j}$ $\rightarrow$ $\cdots$ ($a^{\sigma} \times a^{q})^{n}$\} and thus sets of \{($a^{\sigma} \times a^{q})^{i}$ $\cdots$ ($a^{\sigma} \times a^{q})^{n-x}$\} as opposed to sets of ($a^{i}, ..., a^{n-x}$).  \begin{displayquote} \end{displayquote}
\textbf{Remark 1.5.} Definition 1.4 allows us to note the following: just as programs on arbitrary machines are simply \emph{bit strings}, programs on Ceiling Machines are simply combinations of input (\emph{bits}) and state.  \begin{displayquote} \end{displayquote}
\textbf{Definition 1.6.} We can now offer a more precise definition of a Ceiling Machine. 
Take the 4-tuple $\langle\Ta, A, \La, \delta\rangle$. 
Suppose that this 4-tuple has been \emph{expanded} by the 6-tuple \{$Q, \Gamma, b, \Sigma, q_{0}, A$\} of machine $M$, such that $\La$’ = $\La$ $\cup$ $M$. 
We now write $\langle\Ta, A,$ $\La$$’, \delta\rangle$ for our \textit{CM} which has $\La$’ = $\La$ $\cup$ $M$. 
Adjoin to $\langle\Ta, A,$ $\La$$’, \delta\rangle$ the set $\Gamma$ of tape alphabet symbols such that $A$ $\subseteq$ $\Gamma$, the set $b$ of blank symbols such that $b$ $\in$ $\Gamma$. 
And adjoin also to $\langle\Ta, A,$ $\La$$’, \delta\rangle$ the set $q_{0}$ $\in$ $A$ (the initial state) and the set $F$ $\subseteq$ $A$ (the set of final states). 
We now have:
\begin{displayquote} \begin{center}
	$\langle\Ta, A,$ $\La$$’, \Gamma, \delta\rangle$, 

	where,

	$\delta$: ($A^{\sigma} \times A^{q}$) $\rightarrow$ ($A^{\sigma} \times A^{q}$) $\times$ \{L, R\}. \end{center}

\end{displayquote} 
\textbf{Definition 1.7.} Now we will elaborate a bit on the \emph{specific} programs of a \textit{CM}. 
We can now make use of Martin–L\"of type theory. 
Say that we have a Ceiling Machine $\langle\Ta, A,$ $\La$’, $\delta\rangle$ which has expanded its $\La$ to accommodate a generic 6-tuple machine $M$. 
And say that our \textit{CM} wishes to define some property that $M$ should have, some specific limit on its behavior. 
Suppose we have the following proof-object:
\begin{displayquote} \begin{center}
$\Pi\sigma_{i},\sigma_{j} : P(\sigma_{i},\sigma_{j}).$ $R^{x}(\sigma_{i},\sigma_{j})$ $\Sigma$ $q_{i},q_{j} : Q(q_{i},q_{j}).$ $R^{y}(q_{i},q_{j}) \times ((\delta(q_{i},q_{j}) + \delta(q_{j},q_{i}$)). \end{center}
\end{displayquote}
In everyday language, “for all input symbols $\sigma_{i}$ and $\sigma_{j}$ of the set $P$ of input satisfying some property $R^{x}$ there exist a $q_{i}$ and a $q_{j}$ of the set $Q$ of state satisfying some property $R^{y}$ and behaving such that each $q_{i}$ leads to $q_{j}$ or the converse.”

A \emph{program} leading to this proof-object in a \textit{CM} would simply be a string \{($a^{\sigma}$ $\times$ $a^{q})^{i} \rightarrow (a^{\sigma}$ $\times$ $a^{q})^{j} \rightarrow \cdots (a^{\sigma}$ $\times$ $a^{q})^{n}$\} such that sets of \{($a^{\sigma}$ $\times$ $a^{q})^{i} \cdots (a^{\sigma}$ $\times$ $a^{q})^{n-x}$\} of string \{($a^{\sigma}$ $\times$ $a^{q})^{i} \rightarrow (a^{\sigma}$ $\times$ $a^{q})^{j} \rightarrow \cdots (a^{\sigma} \times a^{q})^{n}$\} correspond to combinations of $j$ and $p$ of some sequence of derivations ($j_{1}, ..., j_{i-1}, p_{i}$) where each $j$ and $p$ of ($j_{1}, ..., j_{i-1}, p_{i}$) is a formula $\varphi$ in \textit{CM}’s $\Ta$.
We can call the proof-object,
\begin{displayquote} \begin{center}
$\Pi\sigma_{i},\sigma_{j} : P(\sigma_{i},\sigma_{j}).$ $R^{x}(\sigma_{i},\sigma_{j})$ $\Sigma$ $q_{i},q_{j} : Q(q_{i},q_{j}).$ $R^{y}(q_{i},q_{j}) \times ((\delta(q_{i},q_{j}) + \delta(q_{j},q_{i}$)), \end{center}
\end{displayquote}
and its (\textbf{$j_{1}$}, ..., \textbf{$j_{i-1}$}, $p_{i}$) an \textit{initialization program} or \textbf{IP} for machine $M$ on \textit{CM}. 
Thus:
\begin{displayquote} \begin{center}
	$\textbf{IP}_{x}$ $\equiv$ $\Pi\sigma_{i},\sigma_{j} : P(\sigma_{i},\sigma_{j}).$ $R^{x}(\sigma_{i},\sigma_{j})$ $\Sigma$ $q_{i},q_{j} : Q(q_{i},q_{j}).$ $R^{y}(q_{i},q_{j}) \times ((\delta(q_{i},q_{j}) + \delta(q_{j},q_{i}$)),

including ($j_{1}, ..., j_{i-1}$). \end{center}

\end{displayquote}
\textbf{Definition 1.8.} Now that we have introduced an example of a so-called initialization program, $\textbf{IP}_{x}$, we can offer another sort of program of a \textit{CM}. 
We will introduce a \emph{schema program} or \textbf{SP}. 
A Ceiling Machine uses an \textbf{SP} to “structure” the behavior of its \emph{embedded machine} $M$ of which its $\La$’ is an expansion $\La$’ = $\La$ $\cup$ $M$. 
We can define the proof-object of a schema program \textbf{SP} as the following:
\begin{displayquote} \begin{center}
	$\Pi$ $\_$:$\Ga$($\_$). $\Sigma$.I($x$, $\_$),

where “$\_$” denotes a variable for a relevant “spot” in the $n$-tuple and corresponding transition function of machine $M$, $\Ga$ is called a schema set and encodes the relevant structure of $M$, “$D$” denotes a type former of proof-objects of declarations (variable assignments in this case), and $x$ is any given declaration of the variable $\_$. \end{center}
\end{displayquote}
An English translation of this is something like, “for every variable \_ of machine $M$’s $n$-tuple structure, there exists a declaration of the variable, $x$.”

Consider a \textit{CM}’s transition function,
\begin{displayquote} \begin{center}
	$\delta$: ($A^{\sigma} \times A^{q}$) $\rightarrow$ ($A^{\sigma} \times A^{q}$) $\times$ \{L, R\}. \end{center}
\end{displayquote}
The program \textbf{SP} is simply an assignment and corresponding transition mapping,
\begin{displayquote} \begin{center}
$\delta$: ($\textbf{A}^{\SIGMA} \times \textbf{A}^{\textbf{q}}) \rightarrow (\textbf{A}^{\SIGMA} \times \textbf{A}^{\textbf{q}}) \times \{$\textbf{L}, R$\}$. \end{center}
\end{displayquote}
of specific $\textbf{A}^{\SIGMA}$ and $\textbf{A}^{\textbf{q}}$ and a corresponding transition between these assigned $\textbf{A}^{\SIGMA}$ and $\textbf{A}^{\textbf{q}}$. 
The program \textbf{SP} is thus not a mere “simulation” of an embedded machine. 
Rather, the programs leading up to \textbf{SP} comprehensively \emph{emulate} (produce) the behavior of the embedded machine. 
The specifics of how this process is implemented are immaterial to the logic of the proof.
\newpage
Graphically, the program \textbf{SP} can be identified with the following blue transitions and states labeled with “$\tau$,” specific programmatic types:

\begin{figure}[h]
    \centering
    \includegraphics[width=\textwidth]{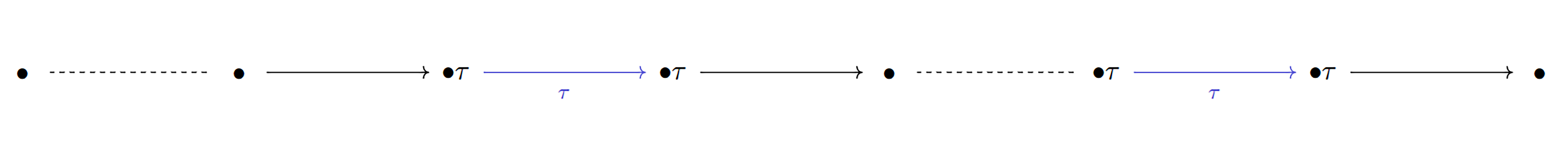}
\end{figure}

Such a schema program \textbf{SP} operates alongside classes of initialization programs \textbf{IP} and another distinguished classes of \textit{CM} program: something we might as well call the schema configuration program or \textbf{$\Ga$CP} for short. The \textbf{$\Ga$CP} ensures that specific initialization programs for \textit{CM}’s embedded machine $M$ get properly “connected to” \emph{programs which lead to} the proof-object of \textbf{SP}:
\begin{displayquote} \begin{center}
$\Pi$ $\_$:$\Ga$($\_$). $\Sigma$.I($x$, $\_$). \end{center}

\end{displayquote}
	\textbf{Remark 1.9.} Where ($j_{1}, ..., j_{i-1}, p_{i}$) of the proof object of \textbf{SP} is stored in a \textit{CM}’s $\Ta$, the actualization or implementation of \textbf{SP} through $A$-transitions corresponds to a program \emph{performing a computation} on the embedded machine $M$. 
It is in this sense that the \textit{CM} is not merely simulating its embedded machine $M$, but it is \emph{literally executing its computations}. 
\textit{CM} is an \emph{emulator} of its embedded machines.
This is a fine point that will be of great importance in our main results. \begin{displayquote} \end{displayquote}
	\textbf{Remark 2.0.} An immediate analogy between \textbf{IP} and \textbf{SP} is that between a high-level programming language and a machine language. 
We might say, \textbf{IP} : \textbf{SP} :: (High-Level Programming Language) : (Machine Language). 
For instance, the proof object,
\begin{displayquote} \begin{center}
$\Pi\sigma_{i},\sigma_{j} : P(\sigma_{i},\sigma_{j}).$ $R^{x}(\sigma_{i},\sigma_{j})$ $\Sigma$ $q_{i},q_{j} : Q(q_{i},q_{j}).$ $R^{y}(q_{i},q_{j}) \times ((\delta(q_{i},q_{j}) + \delta(q_{j},q_{i}$)), \end{center}
\end{displayquote}
is an instruction on \textit{CM}, delimiting the behavior of $M$ (whose output the programs of \textit{CM} ultimately produce). 
Such a proof object as $\textbf{IP}_{x}$, in its current form, obviously cannot be recognized and executed by the machine $M$. 
$\textbf{IP}_{x}$ resides in a “domain” of a higher level than does $M$. 
On the other hand, the proof-object of \textbf{SP},
\begin{displayquote} \begin{center}
$\Pi$ $\_$:$\Ga$($\_$). $\Sigma$.I($x$, $\_$). \end{center}
\end{displayquote}
declares variables of $M$ itself (members of its $n$-tuple definition). 
The precise data that \textbf{SP} declares has a direct impact on the behavior of $M$.

Under this metaphor, the schema configuration program \textbf{$\Ga$CP} serves as the compiler. \textbf{$\Ga$CP} ensures that the proof object,
\begin{displayquote} \begin{center}
$\Pi\sigma_{i},\sigma_{j} : P(\sigma_{i},\sigma_{j}).$ $R^{x}(\sigma_{i},\sigma_{j})$ $\Sigma$ $q_{i},q_{j} : Q(q_{i},q_{j}).$ $R^{y}(q_{i},q_{j}) \times ((\delta(q_{i},q_{j}) + \delta(q_{j},q_{i}$)), \end{center}
\end{displayquote}
“makes it” to the proof-object,
\begin{displayquote} \begin{center}
$\Pi$ $\_$:$\Ga$($\_$). $\Sigma$.I($x$, $\_$). \end{center}
\end{displayquote}
in a format such that (1) the $n$-tuple structure and behavior of $M$ is not impaired and, thus, (2) $M$ “recognizes” this input from \textit{CM}. \begin{displayquote} \end{displayquote}
	\textbf{Definition 2.1.} We can now define the program \textbf{$\Ga$CP}. 
	Take the proof-object,
\begin{displayquote} \begin{center}
$\Pi\sigma_{i},\sigma_{j} : P(\sigma_{i},\sigma_{j}).$ $R^{x}(\sigma_{i},\sigma_{j})$ $\Sigma$ $q_{i},q_{j} : Q(q_{i},q_{j}).$ $R^{y}(q_{i},q_{j}) \times ((\delta(q_{i},q_{j}) + \delta(q_{j},q_{i}$)) \end{center}
\end{displayquote}
Each \emph{type former} in this proof-object ($P, R^{x}, R^{y},$ etc.) is such that it adheres to a binary condition. 
For example, the type former $R^{x}$ could have the following conditions:
\[ \text{$R^{x}$}= \begin{cases}
	\text{\textit{true}}& \text{\emph{if} $\sigma_{i}$ \emph{is such that} $\sigma_{i-1}q\sigma_{i}$ $\Marr$ $\sigma_{i}q\sigma_{j}$}\\
				\text{\textit{false}}& \text{\textit{if otherwise}}
\end{cases} \] 
In this case, the type former $R^{x}$ would be of proof-objects of classes of input symbols $\sigma$ such that
\begin{displayquote} \begin{center}
	$\sigma_{i-1}q\sigma_{i}$ $\Marr$ $\sigma_{i}q\sigma_{j}$ \end{center}
\end{displayquote}
In the case of the type former $Q$ of proof-objects of state symbols, the condition for true would simply look like so:
\begin{displayquote} \begin{center}
	\textit{true if $q_{i}$ and $q_{j}$ are members of $Q$.} \end{center}
\end{displayquote}
These sorts of conditions for type formers comprise the first component of the program \textbf{$\Ga$CP}. 
The second component of \textbf{$\Ga$CP} is related to the specific \emph{assignments} of symbols to each type former. 
For instance, if the condition for $R^{x}$:\emph{true} is met, then we would have something like:
\begin{displayquote} \begin{center}
$\SIGMA_{1}q\SIGMA_{2}$ $\Marr$ $\SIGMA_{2}q\SIGMA_{3}$ \end{center}
\end{displayquote}
with each $\sigma_{n}$ denoting a \emph{specific symbol} from the alphabet $\Sigma$.

The third component of program \textbf{$\Ga$CP} simply ensures that the assignment of specific symbols to each type former accords with the usual notion of Tarskian \emph{satisfaction}. That is, if we had the clause,
\begin{displayquote} \begin{center}
	$\neg\Sigma$ $q_{i},q_{j}$ : $Q(q_{i},q_{j}).$ $R^{y}(q_{i},q_{j})$, \end{center}
\end{displayquote}
\textbf{$\Ga$CP} would ensure that there were no $q_{i},q_{j}$ meeting the condition of $R^{y}$. 
The rest of the nature of this component of \textbf{$\Ga$CP} should be self-explanatory.

The fourth and final component of the program \textbf{$\Ga$CP} simply ensures that the specific symbols as in the condition,
\begin{displayquote} \begin{center}
	$\SIGMA_{1}q\SIGMA_{2}$ $\Marr$ $\SIGMA_{2}q\SIGMA_{3}$
\end{center}
\end{displayquote}
are mapped to the proper portions of the so-called \emph{schema set} of $M$. 
Recall the following proof-object of \textbf{SP}:
\begin{displayquote} \begin{center}
$\Pi$ $\_$:$\Ga$($\_$). $\Sigma$.I($x$, $\_$). \end{center}
\end{displayquote}
A schema set is simply a data structure or a portion thereof containing all the relevant structure of the embedded machine $M$’s behavior (as according to its $n$-tuple definition). 
It matters not the precise implementation of such a schema set. 
All that matters is that the information from initialization programs for $M$, $\textbf{IP}^M$, is properly translated into the proof-object of \textbf{SP}.

According to all four specified components above, the program \textbf{$\Ga$CP} may be defined on the following proof-object (that is, the precise definition of \textbf{$\Ga$CP} is the proof-object $p_{i}$ below, in addition to all corresponding judgements: $j_{1}, ..., j_{i-1}, p_{i}$):
\begin{displayquote} \begin{center}
($\Pi$$\kappa$,$a.\kappa$ :$K.a$:$A.$ $C(\kappa, a))$ $\Rightarrow$ ($\Pi$$d_{i}$, ..., $d_{n}$.$\Ga$(d, $\_$))  

	where $\kappa$ are the conditions of type formers of \textbf{IP}-program proof-objects, $a$ are the specific instances of symbols from $M$’s $n$-tuple definition, $C$ is a type former of proof-objects of \emph{checked} conditions $\kappa$ and instances $a$; $d_{i}, ..., d_{n}$ are declarations of the variable $\_$ in the schema set $\Ga$ of the machine $M$.
	\end{center} 
\end{displayquote}
\textbf{Remark 2.2.} Given the \textit{CM} $\langle\Ta, A, \La$’$, \Gamma, \delta\rangle$ and the generic machine $M$ \{$Q, \Gamma, b, \Sigma, q_{0}, A, \delta$\}, it might not be immediately apparent how the \emph{output} of the two machines is defined. 
But the structure of the output of a \textit{CM} $\langle\Ta, A, \La$’$, \Gamma, \delta\rangle$ is actually quite straightforward. 
All we must do is endow our \textit{CM} with a \emph{program} we may call \textbf{Skip}. 
And, if need be, a similar program could be defined for \emph{input}.

Say, for example, that $M$ has just erased the tape symbol $\sigma_{x}$ and written the symbol $\sigma_{y}$. \textit{CM} then \emph{stores} this symbol $\sigma_{y}$ and its specific square on the tape $s$ in memory. 
Accordingly, when it comes time for \textit{CM} to operate upon tape symbols, its behavior will be constrained by each corresponding tape symbol $\sigma$ and square $s$ it has stored in memory. 
Thus, if \textit{CM} comes across a ($\sigma_{i}$ $|$ $s_{i}$) pair from $M$, it will simply \emph{skip} over to the next available square. 
A precise, inductive definition of such a program should be immediately obvious.	\begin{displayquote} \end{displayquote}
	\textbf{Lemma 2.3.} We will now prove that \emph{any generic} \textit{CM} \emph{is Turing equivalent}. 
This is extremely straightforward, nearly so much as to be trivial. 
But this lemma is crucial, for obvious reasons.

Consider an arbitrary \textit{CM} $\langle\Ta, A, \La$’, $\Gamma, \delta\rangle$, such that $\La$’ = $\La \cup M$ for some arbitrary $M$ \{$Q, \Gamma, b, \Sigma, q_{0}, A$\}. 
Any program that can be run on $M$ (any graph of configurations) can also be run on \textit{CM}, as $\langle\Ta, A, \La$’, $\Gamma, \delta\rangle$ is simply a Turing Machine with a programmatic extension $\langle\Ta, A, \La\rangle$. 
Recall that a \textit{CM} has the following state transition function:
\begin{displayquote} \begin{center}
$\delta$: ($A^{\sigma} \times A^{q}$) $\rightarrow$ ($A^{\sigma} \times A^{q}$) $\times$ \{L, R\}. \end{center}
\end{displayquote}
Thus, any state transition graph $q_{i} \rightarrow q_{j} \cdots \rightarrow q_{n}$ of a generic \emph{deterministic} machine $M$ can be reproduced by a \textit{CM} (a precise definition of a \emph{nondeterministic CM} will be offered shortly). \hfill$\blacksquare$ \begin{displayquote} \end{displayquote}
	\newpage
	\textbf{Definition 2.4.} We will now demarcate two sorts of Ceiling Machines: a \emph{Higher Ceiling Machine} or \emph{HCM} and a \emph{Lower Ceiling Machine} or \emph{LCM}. 
	The former is defined as the following:
\begin{displayquote} \begin{center}
	$\langle\Ta, A, \La$’, $\Gamma, \delta\rangle$,

such that ($\La$’ = $\La \cup M_{i}, ..., M_{n}$) where each $M$ is an $n$-tuple definition corresponding to a \emph{nondeterministic} machine (the relevant changes to this \textit{CM}’s transition $\delta$ will be defined shortly). \end{center} 
\end{displayquote}
And the latter is defined as the 5-tuple:
\begin{displayquote} \begin{center}
	$\langle\Ta, A, \La$’, $\Gamma, \delta\rangle$,

such that ($\La$’ = $\La \cup M_{i}, ..., M_{n}$) where each $M$ is an $n$-tuple definition corresponding to a \emph{deterministic} machine. \end{center} 
\end{displayquote}
\textbf{Remark 2.5.} We now arrive at the motivation behind such phrases as “the canonically nondeterministic machine” and the “canonically deterministic machine.” 
Any \textit{HCM} can be called a \emph{canonically nondeterministic machine}, as each one of its $M$-expansions of ($\La$’ = $\La \cup M_{i}, ..., M_{n}$) is from a nondeterministic machine. 
Likewise, any \textit{LCM} can be called a \emph{canonically deterministic machine}, as each one of its $M$-expansions is from a deterministic machine.

Where \textit{HCM} is a nondeterministic machine, it might not be immediately apparent that we are permitted to endow it with \emph{programs} at all. 
It might be tempting to object to the use of the Curry–Howard correspondence in the context of an \textit{NTM} equivalent. 
However, all we must consider is this: where we are concerned only with the sets \{($a^{\sigma} \times a^{q})^{i}$ $\cdots$ ($a^{i} \times a^{q})^{n-x}$\} of a \textit{CM}—recall Definition 1.4—which correspond to formulae $\varphi_{1}$, ..., $\varphi_{n}$ in its $\Ta$, nondeterminism does not pose a threat to our maxim of machines as programs. 
In the opening section of the paper, we mentioned that we need not know what hypothetical programs on a \textit{NTM} “look like.” 
Rather, we can apply a sort of \emph{formal software structuralism} and consider solely the logical rules (proof structure) which the programs conform to. 
In the case of a \textit{NTM}, all we must do to have a program corresponding to one on a \textit{DTM} is have \emph{some instance} along the graph of transition relation of computation sets \{($a^{\sigma} \times a^{q})^{i}$ $\cdots$ ($a^{\sigma} \times a^{q})^{n-x}$\} corresponding to the same (concordantly programmed) formulae $\varphi_{1}, ..., \varphi_{n}$ in the \textit{DTM}’s $\Ta$. 
The only difference with an \textit{NTM} is that it can run multiple programs at \emph{the same time}. \begin{displayquote} \end{displayquote}
	\textbf{Proposition 2.6.} Take any generic \textit{HCM} $\langle\Ta, A, \La$’, $\Gamma, \delta\rangle$ with arbitrary expansion ($\La$’ = $\La \cup M_{i}, ..., M_{n}$). 
Single out any specific $M$ of the set \{$M_{i}, ..., M_{n}$\} of \textit{HCM}’s expansion. 
Recall the following: 

We can call the proof-object,
\begin{displayquote} \begin{center}
$\Pi\sigma_{i},\sigma_{j} : P(\sigma_{i},\sigma_{j}).$ $R^{x}(\sigma_{i},\sigma_{j})$ $\Sigma$ $q_{i},q_{j} : Q(q_{i},q_{j}).$ $R^{y}(q_{i},q_{j}) \times ((\delta(q_{i},q_{j}) + \delta(q_{j},q_{i}$)), \end{center}
\end{displayquote}
and its (\textbf{$j_{1}$}, ..., \textbf{$j_{i-1}$}, $p_{i}$) an \textit{initialization program} or \textbf{IP} for machine $M$ on \textit{CM}. Thus:
\begin{displayquote} \begin{center}
	$\textbf{IP}_{x}$ $\equiv$ $\Pi\sigma_{i},\sigma_{j} : P(\sigma_{i},\sigma_{j}).$ $R^{x}(\sigma_{i},\sigma_{j})$ $\Sigma$ $q_{i},q_{j} : Q(q_{i},q_{j}).$ $R^{y}(q_{i},q_{j}) \times ((\delta(q_{i},q_{j}) + \delta(q_{j},q_{i}$)),

including ($j_{1}, ..., j_{i-1}$).  \end{center}
\end{displayquote}
Fashion a class \texttt{IP} consisting of arbitrary initialization programs like $\textbf{IP}_{x}$ above.
Or, to simplify things, we may simply consider that $\textbf{IP}_{x}$ is the only initialization program (assuming that the conditions associated with each type former hold). 
We might say that, outside the specific initialization programs fed to $M$ by \textit{HCM}, that $M$ can be left to “make its own choice” through its finite control, beyond adhering to the members of \texttt{IP}. 
In this case, it would be easy enough to adjoin to \textit{HCM} the relevant program specifying as much. 
For now, however, consider an arbitrary class \texttt{IP}.

Recall that the \textbf{$\Ga$CP} program is defined like so:
\begin{displayquote} \begin{center}
($\Pi$$\kappa$,$a.\kappa$ :$K.a$:$A.$ $C(\kappa, a))$ $\Rightarrow$ ($\Pi$$d_{i}$, ..., $d_{n}$.$\Ga$(d, $\_$)),

including ($j_{1}, ..., j_{i-1}$).\end{center}
\end{displayquote}
Also recall that the proof-object of \textbf{SP} is defined as the following:
\begin{displayquote} \begin{center}
$\Pi$ $\_$:$\Ga$($\_$). $\Sigma$.I($x$, $\_$). \end{center}
\end{displayquote}
Now associate with a specific $M$ chosen from the set \{$M_{i}, ..., M_{n}$\} of \emph{HCM}’s expansion a class \texttt{IP}. 
Then, consider that the program \textbf{$\Ga$CP} ensures that the data from \texttt{IP} delivers to the proof object of \textbf{SP} declared variables in $\Ga$.

Insert a strict partial order into the class \texttt{IP}, such that \texttt{IP} = \{$\textbf{IP}_{1} < \textbf{IP}_{2} < ... < \textbf{IP}_{n}$\}. 
Then fashion a string \{($a^{\sigma} \times a^{q})^{i} \times (a^{\sigma} \times a^{q})^{j} \times \cdots (a^{\sigma} \times a^{q})^{n}$\} of \textit{HCM} such that for each \textbf{IP} of the set \texttt{IP} there is a corresponding set from \{($a^{\sigma} \times a^{q})^{i} \cdots (a^{\sigma} \times a^{q})^{n-x}$\} which corresponds to a sequence of judgements ($j_{1}, ..., j_{i-1}, p_{i}$) where each $j$ and $p$ of ($j_{1}, ..., j_{i-1}, p_{i}$) is a formula $\varphi$ in \textit{HCM}’s $\Ta$. 
Also allow room for any intermediary programs of the set \texttt{IP} = \{$\textbf{IP}_{1} < \textbf{IP}_{2} < ... < \textbf{IP}_{n}$\}. 
In other words, wherever there are $x$ such that ($x < \textbf{IP}$) or ($\textbf{IP} < x$), ensure that HCM’s string \{($a^{\sigma} \times a^{q})^{i} \times (a^{\sigma} \times a^{q})^{j} \times \cdots (a^{\sigma} \times a^{q})^{n}$\} accommodates for such $x$.

Call this string corresponding to the set \texttt{IP}, with any intermediary programs or computations $x$, \texttt{String IP}. 
Then, we may adjoin \texttt{String IP} to the program \textbf{$\Ga$CP} taking all \textbf{IP} to $\Pi$ $\_$:$\Ga$($\_$). $\Sigma.I(x,$ $\_$).

To do so, we can start by fashioning a set \{($a^{\sigma} \times a^{q})^{i} \cdots (a^{\sigma} \times a^{q})^{n-x}$\} of \textit{HCM} corresponding to:
\begin{displayquote} \begin{center}
($\Pi$$\kappa$,$a.\kappa$ :$K.a$:$A.$ $C(\kappa, a))$ $\Rightarrow$ ($\Pi$$d_{i}$, ..., $d_{n}$.$\Ga$(d, $\_$)),

	including ($j_{1}, ..., j_{i-1}$),

where ($j_{1}, ..., j_{i-1}$) excludes any ($j_{1}, ..., j_{i-1}, p_{i}$) already expressed in \textit{HCM}’s \texttt{String IP}, which need not be expressed in ($j_{1}, ..., j_{i-1}, p_{i}$) of \textbf{$\Ga$CP}. \end{center}
\end{displayquote}
Call this string \texttt{String $\Ga$CP}.

In addition, we can also fashion the set  \{($a^{\sigma} \times a^{q})^{i} \cdots (a^{\sigma} \times a^{q})^{n-x}$\} of \textit{HCM} corresponding to,
\begin{displayquote} \begin{center}
	$\Pi$ $\_$:$\Ga$($\_$). $\Sigma.I(x,$ $\_$),

	including ($j_{1}, ..., j_{i-1}$),

where ($j_{1}, ..., j_{i-1}$) excludes any ($j_{1}, ..., j_{i-1}, p_{i}$) already expressed in \textit{HCM}’s \texttt{String IP} or \texttt{String $\Ga$CP} which need not be expressed in ($j_{1}, ..., j_{i-1}, p_{i}$) of \textbf{SP}. \end{center}
\end{displayquote}
Call this string \texttt{String SP}.

We can then form the string \{\{\texttt{String IP}\} $\cup$ \{\texttt{String $\Ga$CP}\} $\cup$ \{\texttt{String SP}\}\}. \hfill$\blacksquare$ \begin{displayquote} \end{displayquote}
\textbf{Remark 2.7.} Proposition 2.6 allows us to now introduce an important phrase from the abstract—\textit{CNDS} or the \emph{canonically nondeterministic solution}. 
In Remark 2.5 we mentioned that any arbitrary \textit{HCM} can be called a canonically nondeterministic machine, as its expansion ($\La$’ = $\La \cup M_{i}, ..., M_{n}$) is such that each $M$ is nondeterministic. 
Accordingly, we may term the string \{\{\texttt{String IP}\} $\cup$ \{\texttt{String $\Ga$CP}\} $\cup$ \{\texttt{String SP}\}\} the canonically nondeterministic solution or \textit{CNDS} (whose accompanying \textit{problem} will be defined shortly). \begin{displayquote} \end{displayquote}
\textbf{Lemma 2.8.} Expanding on Remark 2.7, we will now prove that \textit{CNDS} is produced by a nondeterministic Turing Machine. 
Lemma 2.3 shows us that any arbitrary \textit{CM} (\textit{LCM} or \textit{HCM}) is already Turing equivalent, as any \textit{CM} is just a \textit{TM} with the programmatic extension $\langle\Ta, A, \La\rangle$. 
Remember that any \textit{CM} has a transition function identical to that of a \textit{TM}. 
However, we only previously offered a transition function for an LCM or a canonically \textit{deterministic} machine:
\begin{displayquote} \begin{center}
	$\delta$: ($A^{\sigma} \times A^{q}) \rightarrow (A^{\sigma} \times A^{q}) \times \{$L, R$\}$. \end{center}
\end{displayquote}
It is easy enough to fashion a transition relation for an arbitrary HCM:
\begin{displayquote} \begin{center}
	$\delta$: ($A^{\sigma} \times A^{q}) \times (A^{\sigma} \times A^{q}) \times \{$L, R$\}$. \end{center}
\end{displayquote}
Accordingly, this makes the nondeterministic version of the schema program SP simply look like so:
\begin{displayquote} \begin{center}
$\delta$: ($\textbf{A}^{\SIGMA} \times \textbf{A}^{\textbf{q}}) \times (\textbf{A}^{\SIGMA} \times \textbf{A}^{\textbf{q}}) \times \{$L, \textbf{R}$\}$. \end{center}
\end{displayquote}
an assignment of specific instances, a movement of the head and a corresponding transition relation between the assigned instances.
\newpage
Graphically, nondeterministic instances of \textbf{SP} can be identified with the following blue transitions and states labeled with “$\tau$,” specific programmatic types:

\begin{figure}[h]
    \centering
    \includegraphics[width=4in]{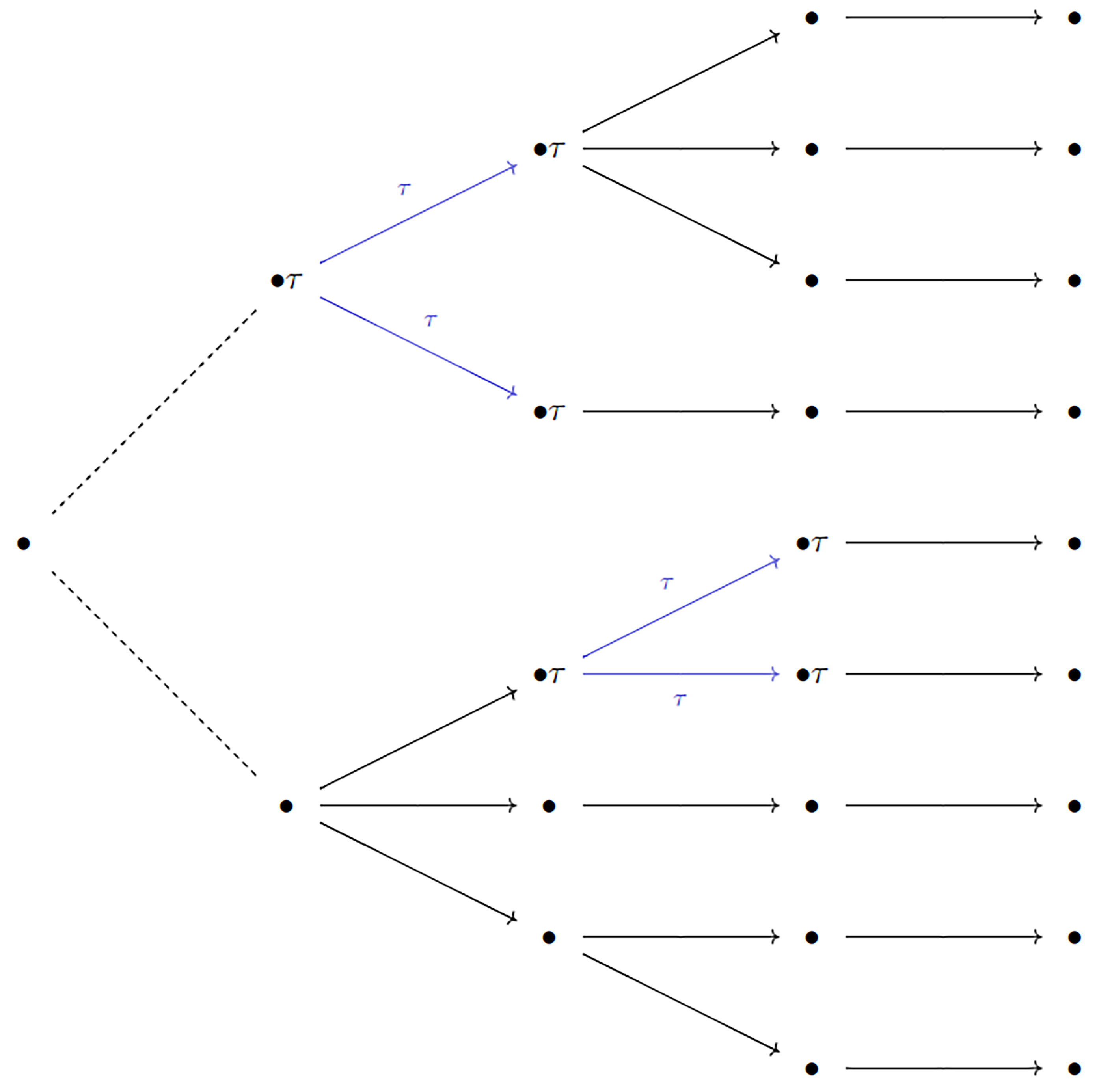}
\end{figure}

Thus, it becomes immediately clear that a \textit{HCM} is just a \emph{nondeterministic TM} with the extension $\langle\Ta, A, \La\rangle$, which is solely built from the transition relation  = ($A^{\sigma} \times A^{q}) \times (A^{\sigma} \times A^{q}) \times \{$L, R$\}$. \textit{CNDS} is thus produced by a \textit{NTM} equivalent. \hfill$\blacksquare$ \begin{displayquote} \end{displayquote}
\textbf{Remark 2.9.} We are now nearly ready for the main results. 
I will now briefly outline the coming steps. 
First, we must define a suitable problem whose answer is the solution \textit{CNDS}. 
We must find a language $L$ which, when fed into a \textit{CM}, would allow this \textit{CM} to follow the state transition path corresponding to \textit{CNDS}. 
The obvious answer: fashion a problem which is an injunction to \textit{replicate CNDS}. 
In this manner, if a \textit{CM} accepts the language $L$, the state transitions of \textit{CNDS} would be found somewhere in the behavior of \textit{CM}.

It matters not whether the \textit{CM} would be able to replicate the transition path verbatim. 
Rather, it would suffice if each set \{($a^{\sigma} \times a^{q})^{i} \cdots (a^{\sigma} \times a^{q})^{n-x}$\} were reproduced in \textit{CM}, excluding any intermediary transitions. 
However, we will require not only a reproduction of each set \{($a^{\sigma} \times a^{q})^{i} \cdots (a^{\sigma} \times a^{q})^{n-x}$\}, but that each reproduced set also corresponds to \emph{the same members} of $\Ta$ and $\La$ in CM’s memory. 
In what follows, \emph{we will assume this criterion}. \begin{displayquote} \end{displayquote}
	\textbf{Definition 3.0} We will now fashion a specific problem corresponding to the solution that is \textit{CNDS}.
	Call this problem the \textit{canonically nondeterministic problem} or \textit{CNDP}. We will define \textit{CNDP} on the sets \{($a^{\sigma} \times a^{q})^{i} \cdots (a^{\sigma} \times a^{q})^{n-x}$\} of \textit{CNDS}. 
	The gist of \textit{CNDP} is that it necessitates an algorithm \emph{searching} through \textit{LCM}’s memory for instances of \{($a^{\sigma} \times a^{q})^{i} \cdots (a^{\sigma} \times a^{q})^{n-x}$\} corresponding to those from \textit{CNDS} of \textit{HCM}. 
	\textit{CNDP} searches \textit{LCM}’s memory for such instances, and if for each set comprising ($a^{\sigma} \times a^{q}$) of \textit{CNDS} there exists a corresponding set comprising ($a^{\sigma} \times a^{q}$) in \textit{HCM}, \textit{CNDP}'s corresponding algorithm tells the \textit{LCM} to \emph{execute} or \emph{run} each corresponding ($a^{\sigma} \times a^{q}$) of the set \{($a^{\sigma} \times a^{q})^{i} \cdots (a^{\sigma} \times a^{q})^{n-x}$\} corresponding to \textit{CNDS}. 
	If the Lower Ceiling Machine does indeed have each corresponding set in its memory, the behavior of \textit{LCM} is then such that \{($a^{\sigma} \times a^{q})^{i} \cdots (a^{\sigma} \times a^{q})^{n-x}$\} of \textit{CNDS} are produced, alongside arbitrary intermediary sets, where (($a^{\sigma} \times a^{q}) < x$) or ($x < (a^{\sigma} \times a^{q}$)), where “$<$” denotes a partial order corresponding to time of execution. 
	A solution to \textit{CNDP} for an \textit{LCM} would thus be the following:
\begin{displayquote} \begin{center}
\{($a^{\sigma} \times a^{q})^{i} \cdots (a^{\sigma} \times a^{q})^{n-x}$\},

with the accompanying intermediary sets $x$.	\end{center}
\end{displayquote}
Let us now define the algorithm corresponding to \textit{CNDP}. 
Define said algorithm as the following proof-object in conjunction with the relevant judgements ($j_{1}, ..., j_{i-1}$):
\begin{displayquote} \begin{center}
($\Pi(a^{\sigma} \times a^{q})^{<}:$\textit{CNDS.S}$(a^{\sigma} \times a^{q})^{<}.$ $\Sigma$ ($a^{\sigma} \times a^{q}):$\textit{LCM.I}$((a^{\sigma} \times a^{q}), (a^{\sigma} \times a^{q}))$ \\ $\Rightarrow$ ($\Pi(a^{\sigma} \times a^{q})^{<}:$\textit{LCM.E}($a^{\sigma} \times a^{q})^{<}$),

	where $S$ is a type former of proof-objects of \emph{search} statements, $I$ is a type former of proof-objects of \emph{identity} statements (\textit{equality} between stored data), $E$ is a type former of proof-objects of \textit{execute} or \emph{run} statements, and the superscript $<$ is the partial order on sets ($a^{\sigma} \times a^{q}$); we somewhat sloppily use “($a^{\sigma} \times a^{q}$)” to denote arbitrary sets of \{($a^{\sigma} \times a^{q})^{i} \cdots (a^{\sigma} \times a^{q})^{n-x}$\}.	\end{center}
\end{displayquote}
\textbf{Proof 3.1.} We can now prove that P $\neq$ NP.
Take any arbitrary \textit{HCM} $\langle\Ta, A, \La$’, $\Gamma, \delta\rangle^{HCM}$ and any arbitrary \textit{LCM} $\langle\Ta, A, \La$’, $\Gamma, \delta\rangle^{LCM}$. 
Then take the string \{\{\texttt{String IP}\} $<$ \texttt{String $\Ga$CP}\} $<$ \texttt{String SP}\}\} produced by \textit{HCM}. 
We call this string the \textit{CNDS}. 
Lemma 2.3. shows that this string, \textit{CNDS}, has been produced by a nondeterministic Turing machine. Lemma 2.3. also shows that \textit{LCM} is equivalent to a deterministic Turing machine.

Thus, if we can give \textit{LCM} a language $L$ that it \emph{cannot replicate} (or a program which it cannot execute!), that \textit{HCM can replicate}, we can offer an instance of a problem solvable by a nondeterministic machine, yet unsolvable by a deterministic one. 
Such a problem would ensure that P $\neq$ NP.

Take the algorithm corresponding to \textit{CNDP}, defined above as the following proof-object with relevant judgements ($j_{1}, ..., j_{i-1}$):
\begin{displayquote} \begin{center}
($\Pi(a^{\sigma} \times a^{q})^{<}:$\textit{CNDS.S}$(a^{\sigma} \times a^{q})^{<}.$ $\Sigma$ ($a^{\sigma} \times a^{q}):$\textit{LCM.I}$((a^{\sigma} \times a^{q}), (a^{\sigma} \times a^{q}))$ \\ $\Rightarrow$ ($\Pi(a^{\sigma} \times a^{q})^{<}:$\textit{LCM.E}($a^{\sigma} \times a^{q})^{<}$),

where for each set of computations \{($a^{\sigma} \times a^{q})^{i}$ $\cdots$ ($a^{\sigma} \times a^{q})^{n-x}$\} corresponding to a program of \textit{CNDS} (which is a proof, ($j_{1}, ..., j_{i-1}, p_{i}$)), \textit{LCM} is such that the same computation sets \{($a^{\sigma} \times a^{q})^{i}$ $\cdots$ ($a^{\sigma} \times a^{q})^{n-x}$\} lead to the same ($j_{1}, ..., j_{i-1}, p_{i}$) in \textit{LCM}’s $\Ta$ as they do in \textit{HCM}’s $\Ta$ (and thus \textit{CNDS}). \end{center}
\end{displayquote}
We will call this algorithm \textbf{REP}. 
We have spoken rather loosely about \textit{CNDS} and its corresponding algorithm \textbf{REP} of \textit{CNDP} above, and there is one small addendum. 
I have neglected something quite obvious: in order for our \textit{LCM} to run the \textbf{REP} algorithm, it must have already been fed the input corresponding to the string \textit{CNDS}. 
Thus, we must feed the string \{($a^{\sigma} \times a^{q})^{i}$ $\cdots$ ($a^{\sigma} \times a^{q})^{n-x}$\} (where each symbol is fashioned into a tape symbol $\sigma$ for input) corresponding to \textit{CNDS} into our \textit{LCM} before it can run the algorithm \textbf{REP}.\footnote{More precisely, we can fashion an input $\textit{CNDS}^{Q}$, a decision problem which asks whether or not \textit{CNDS} can be replicated. 
All this does is eliminate the final execution (replication) clause from the algorithm \textbf{REP}. 
In other words, if all corresponding sets could be found in memory, the query $\textit{CNDS}^{Q}$ would be answered. 
We will use this version of \textit{CNDS} for our final results.} 
Accordingly, we define the input into our \textit{LCM}—before the execution of \textbf{REP}—as the following: \{\{\texttt{String IP}\} $<$ \{\texttt{String $\Ga$CP}\} $<$ \{\texttt{String SP}\}\} in hopes that it will produce the transitions (via the algorithm \textbf{REP}), \{($a^{\sigma} \times a^{q})^{i}$ $\cdots$ ($a^{\sigma} \times a^{q})^{n-x}$\}, with accompanying intermediary sets $x$, such that the complement of all said intermediary sets is equivalent to the string that is \textit{CNDS}.\footnote{
	Of course, if we dropped the requirement that each \emph{subprogram} (or set ($j_{1}, ..., j_{i-1}, p_{i}$) of \textit{formulae} in a \textit{CM}’s $\Ta$) of \textit{CNDS} have the same corresponding computation sets \{($a^{\sigma} \times a^{q})^{i} < \cdots < (a^{\sigma} \times a^{q})^{n-x}$\}, all we would need to do is ask our \textit{LCM} to replicate the set \{($j_{1}, ..., j_{i-1}, p_{i})^{i} < \cdots < (j_{1}, ..., j_{i-1}, p_{i})^{n-x}$\} of proofs/programs encoded in \textit{HCM}’s memory. 
	It would be vacuous, in such a case, to ask an \textit{LCM} to replicate \emph{only} the corresponding computation sets.}

More simply, suppose that we give an arbitrary \textit{LCM} the string  \{\{\texttt{String IP}\} $<$ \{\texttt{String $\Ga$CP}\} $<$ \{\texttt{String SP}\}\} plus the \textit{query} “Can this string be replicated on this machine?” or “Does the algorithm \textbf{REP} terminate in the accepting state?” as input. 
We then implement the algorithm \textbf{REP} on \textit{LCM}, in hopes that it can reproduce the input string of \textit{CNDS} as state transition (computation) and give an affirmative answer to the query. 
In other words, we hope that the algorithm \textbf{REP} can do its job.

Suppose that \textit{LCM} has only one machine $M$ in its expansion, such that ($\La$’ = $\La \cup M$).

Recall that leading up to the proof object of \textbf{SP}, $\Pi$ $\_$:$\Ga$($\_$). $\Sigma.I(x,$ $\_$), we have the following program:
\begin{displayquote} \begin{center}
	($\Pi$$\kappa$,$a.\kappa$ :$K.a$:$A.$ $C(\kappa, a))$ $\Rightarrow$ ($\Pi$$d_{i}$, ..., $d_{n}$.$\Ga$(d, $\_$)),	

	including ($j_{1}, ..., j_{i-1}$). \end{center}
\end{displayquote}
We have referred to this program as \textbf{$\Ga$CP}.

Also recall that leading up to the program \textbf{$\Ga$CP} we have various initialization programs, comprising the class \texttt{IP} = \{$\textbf{IP}_{1} < \textbf{IP}_{2} < ... < \textbf{IP}_{n}$\}. 
In order to more straightforwardly deduce the P $\neq$ NP result, we can endow our \textit{HCM} with a particular member of \texttt{IP}. 
We will call this member the \emph{canonical initialization} program of its sole embedded \textit{nondeterministic} $M$, symbolized as $\textbf{IP}_{C}$.
We will define $\textbf{IP}_{C}$ like so:
\begin{displayquote} \begin{center}
	$\Sigma$$q_{i}$,$q_{j}$:$Q(q_{i},q_{j}).$ $\Sigma\delta$:$S(\delta)$:$(\delta(q_{i},q_{j}) \times \delta(q_{x},q_{j}$),

	including ($j_{1}, ..., j_{i-1}$),

where $Q$ is the type former of proof-objects of state statements, $S$ is the type former of proof-objects of transition relation statements, and $q_{x}$ is an arbitrary state also transitioning to $q_{j}$ (in addition to $q_{i}$). \end{center}
\end{displayquote}
We account for $\textbf{IP}_{C}$ in the string \{\{\texttt{String IP}\} $\cup$ \{\texttt{String $\Ga$CP}\} $\cup$ \{\texttt{String SP}\}\}, such that $\textbf{IP}_{C}$ $\in$ \{\texttt{String IP}\}. 
In the context of $\textbf{IP}_{C}$ $\in$ \{\texttt{String IP}\}, the program \textbf{$\Ga$CP} ensures that all members of \texttt{IP} “make it” to the proof object of \textbf{SP}, $\Pi$ $\_$:$\Ga$($\_$). $\Sigma.I(x,$ $\_$). 
Recall that \textit{CNDS} is a computation or string produced by a \textit{nondeterministic machine equivalent}.

Suppose that an \textit{LCM} were fed the input corresponding to the decision problem $\textit{CNDS}^{Q}$ , i.e., the question “can the language corresponding to \textit{CNDS} be \emph{replicated} as a computation on \emph{LCM}?” To answer this query, \textit{LCM} runs the algorithm \textbf{REP} (the version of \textbf{REP} corresponding to the query $\textit{CNDS}^{Q}$),
\begin{displayquote} \begin{center}
	($\Pi(a^{\sigma} \times a^{q})^{<}$:\textit{CNDS.S}$(a^{\sigma} \times a^{q})^{<}$. $\Sigma$ ($a^{\sigma} \times a^{q}$):\textit{LCM.I}(($a^{\sigma} \times a^{q}$), ($a^{\sigma} \times a^{q}$)),

where for each set of computations \{($a^{\sigma} \times a^{q})^{i} \cdots (a^{\sigma} \times a^{q})^{n-x}$\} corresponding to a program of \textit{CNDS} (which is a proof, ($j_{1}, ..., j_{i-1}, p_{i}$)), \textit{LCM} is such that the same computation sets \{($a^{\sigma} \times a^{q})^{i} \cdots (a^{\sigma} \times a^{q})^{n-x}$\} lead to the same ($j_{1}, ..., j_{i-1}, p_{i}$) in \textit{LCM}’s $\Ta$ as they do in \textit{HCM}’s $\Ta$ (and thus \textit{CNDS}), \end{center}
\end{displayquote}
which searches the memory of \textit{LCM} for programs corresponding to formulae (so-called \emph{program templates}) $(\varphi{i}, ..., \varphi{n}$) in its theory $\Ta$.

However, \textit{LCM} \emph{cannot possibly} run the algorithm \textbf{REP} to the solution above. 
To see why this is the case we need only consider the proof object $\Pi$ $\_$:$\Ga$($\_$). $\Sigma.I(x,$ $\_$) of \textit{LCM}. 
Since our \textit{LCM}’s expansion consists of only one \emph{deterministic} machine, its corresponding species set $\Ga$ is tailored to the data structure of $n$-tuple definitions of \emph{deterministic} machines. 
Accordingly, it is impossible that the proof-object,
\begin{displayquote} \begin{center}
	$\Sigma$$q_{i}$,$q_{j}$:$Q(q_{i},q_{j}).$ $\Sigma\delta$:$S(\delta)$:$(\delta(q_{i},q_{j}) \times \delta(q_{x},q_{j}$), \end{center}
\end{displayquote}
exists as an executable program along the computation path to the proof-object, $\Pi$ $\_$:$\Ga$($\_$). $\Sigma.I(x,$ $\_$), for \textit{LCM}. 
There are no proof-objects of transition relation statements “$S$” for \textit{LCM} because of the definition of \textit{LCM}:
\begin{displayquote} \begin{center}
	$\langle\Ta, A, \La$’, $\Gamma, \delta\rangle$,

such that ($\La$’ = $\La \cup M_{i}, ..., M_{n}$) where each $M$ is an $n$-tuple definition corresponding to a \emph{deterministic} machine \emph{and} such that \textit{CM}’s $\delta$ is a \emph{function}. \end{center}
\end{displayquote}
The language $\La$’ of \textit{LCM} is such that there is \emph{no} transition \emph{relation} symbol (there is only the \emph{transition} function of the \textit{LCM} itself!). 
Accordingly, there is no group of formulae $\varphi_{1}, ..., \varphi_{n}$ in its theory $\Ta$ corresponding to a program of its implementation (and, if there were, it would be \emph{vacuous}, \emph{unable to execute the program} \textbf{SP}). 
This is a sort of \emph{programmatic diagonalization} which, clearly, \emph{does not relativize}.

Recall briefly that the \emph{nondeterministic} (\textit{HCM}’s) version of the program \textbf{SP} looks like so:
\begin{displayquote} \begin{center}
$\delta$: ($\textbf{A}^{\SIGMA} \times \textbf{A}^{\textbf{q}}) \times (\textbf{A}^{\SIGMA} \times \textbf{A}^{\textbf{q}}) \times \{$L, \textbf{R}$\}$. \end{center}
\end{displayquote}
Also recall that the \textit{deterministic} (\textit{LCM}’s) version of \textbf{SP} looks like:
\begin{displayquote} \begin{center}
$\delta$: ($\textbf{A}^{\SIGMA} \times \textbf{A}^{\textbf{q}}) \rightarrow (\textbf{A}^{\SIGMA} \times \textbf{A}^{\textbf{q}}) \times \{$\textbf{L}, R$\}$. \end{center}
\end{displayquote}
From this vantage it is exceedingly clear that any attempt by an \textit{LCM} traverse \textit{CNDS} and thus execute the program \textbf{SP} begets a \emph{type error}.

\newpage$ $For more clarity, consider the following graphics:
\vspace*{\fill}
\begin{figure}[h!]
	\centering
	\captionsetup{justification=centering, margin=1cm}
	\includegraphics[width=5in]{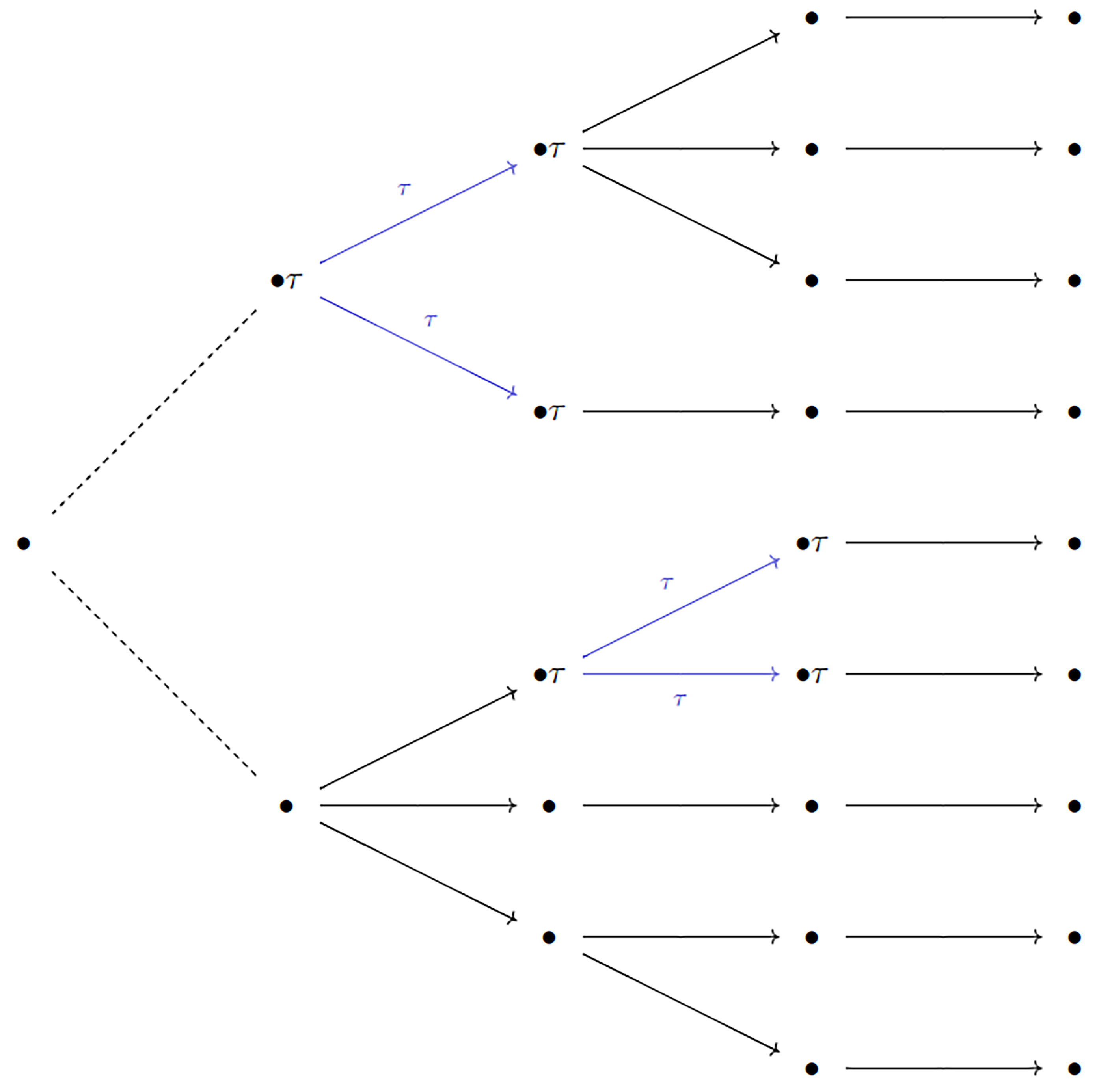}
	\caption{This is a potential tree of state relations of an \textit{HCM}. 
	The blue arrows represent executions of the program \textbf{SP}. 
	The states and accompanying transition relation at the given executions are each labeled with a “$\tau$,” denoting that each has a corresponding type $\tau_{i}$ in the programs of \textit{HCM}. 
	Where the program \textbf{SP} is defined on \textit{both} the states and the \emph{transition itself}, an \textit{LCM} wishing to replicate \textit{CNDS} would have to not only replicate the states, $q_{i}, q_{j}$, themselves but programmatically replicate their corresponding types, $\tau_{i}$, $\tau_{j}$, and the type of the transition relation itself, $\tau_{j-1}$, all in an \emph{executable program}. 
	The term “embedded machine” takes on a pictorial character above, where each instance of \textbf{SP} is identified with a transition of an embedded machine $M$. 
	Whatever the particular ingredients are, for any given $M$, it is embedded in the transitions of its \textit{CM}. 
	Additionally, since a \textit{CM} is a general-purpose programmatic extension of each of its $M$’s, any specifics of each $M$ can be incorporated into its programming.}
\end{figure}
\vspace*{\fill}
\begin{figure}[h!]
	\centering
	\captionsetup{justification=centering, margin=1cm}
	\includegraphics[width=\textwidth]{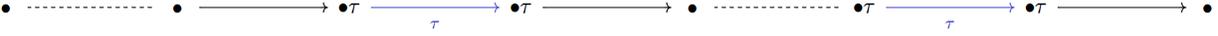}
	\caption{This is a potential graph of state transition of an \textit{LCM}. 
	As above, each blue transition and its accompanying states labeled with “$\tau$” represent an execution of the program \textbf{SP}. 
	Regardless of the specific embedded machine $M$ that an \textit{LCM} emulates, its computational path is \emph{deterministic}. 
	This entails that each transition type $\tau$ (blue path) corresponds to a deterministic, \emph{function} type.}
\end{figure}

\newpage
We have offered an example of a language \textit{CNDS} which \emph{can} be trivially solved by a nondeterministic machine (as \textit{HCM} is the one which produced \emph{CNDS} and can thus trivially reproduce it) but cannot be solved by a deterministic machine (as \textit{LCM} cannot possess the requisite structure in its extension $\langle\Ta, A, \La\rangle$ to enable it to execute the replication algorithm \textbf{REP}). 
Moreover, the emulation of an arbitrary \textit{NTM} by an \textit{HCM}—as found in \textit{CNDS}—is clearly a polynomial-time algorithm. 
\textit{HCM} is endowed with a pre-specified template of $M$’s $n$-tuple structure, and it must simply match the relevant variables with the relevant places in this structure. 
As far as the program \textbf{$\Ga$CP} is concerned, all our \textit{HCM} must do is ensure that these assignments adhere to the usual notions of satisfaction.
This amounts to an elementary constraint satisfaction problem. 
And, since we are permitted to define \textit{CNDS} on only a handful of \texttt{IP} proof-objects, such a procedure is ensured to be in $P$.

Although it \emph{is} the case that a \textit{DTM} can \emph{simulate} an \textit{NTM}, it is \emph{not} the case that a \textit{DTM} can \emph{execute} the computation structure (transition of state via \emph{relation}) of an \textit{NTM}. 
And, as we have defined the algorithm \textbf{REP} on the eventual \emph{programmatic execution} or \emph{emulation} of a given embedded machine (as the proof object of \textbf{SP})—as opposed to a mere \emph{simulation} of it—an \textit{LCM} cannot affirmatively answer the query $\textit{CNDS}^{Q}$.

Even more succinctly: no comprehensive (applicable to all possible \textit{DTM}s) deterministic machine can follow the programmatic path of a nondeterministic machine emulator, as it computes a state transition for one of its emulated machines. 

This much suffices to prove that P $\neq$ NP. \hfill$\blacksquare$ \begin{displayquote} \end{displayquote}
	\textbf{Remark 3.2.} Now we will address a few possible objections to our result. 
When we say that our \textit{LCM} is a so-called canonical example of a deterministic machine, this entails that \emph{no possible embedded deterministic machine can be fashioned to affirmatively answer} $\textit{CNDS}^{Q}$. Remember that each embedded machine $M$ of any arbitrary \textit{LCM} has \textit{LCM}’s transition function \emph{as its own} (see Definition 1.2). And as far as deterministic machines are \emph{classically} conceived—without an accompanying \textit{CM})—it is obvious that they are not equipped with the programmatic structure to solve \textit{CNDS}. 
But there is another, more potent possible sticking point: one may wonder if we could \emph{break the definition of a Lower Ceiling Machine} and endow one of its embedded machines $M$ with a transition \emph{relation} (in hopes to enable the \textit{LCM} to affirmatively answer the query $\textit{CNDS}^{Q}$). We will address this objection below. \begin{displayquote} \end{displayquote}
	\textbf{Remark 3.3.} Consider that we had some embedded \emph{nondeterministic} machine $M$ of a generic \textit{LCM} (going against the definitions we have utilized). 
Say that we endowed the embedded machine $M$ with \emph{its own} transition \emph{relation} $\delta$, such that $M$ = \{$Q, \Gamma, b, \Sigma, q_{0}, A, \delta$\} (again going against previous definitions). 
Thus, we would try to prove that a deterministic machine (the \textit{LCM}) could run along the computation path of \textit{CNDS}, to the transition relation of $M$. 
However, this would be a problem. 
Recall what was said in Definition 1.2—“The transition function/relation in the extension $\langle\Ta, A, \La, \delta\rangle$ of a \textit{CM} serves as the governing function/relation for each embedded machine $M$ of \textit{CM}.” 
As a result of this, the way the program \textbf{SP} is defined—the program which carries out the computations of the embedded $M$—is such that \emph{even if} we had some \textit{LCM} able to execute the program $\textbf{IP}_{C}$ (as would trivially be the case with an embedded \textit{NTM} of an arbitrary \textit{LCM}),
\begin{displayquote} \begin{center}
	$\Sigma$$q_{i}$,$q_{j}$:$Q(q_{i},q_{j}).$ $\Sigma\delta$:$S(\delta)$:$(\delta(q_{i},q_{j}) \times \delta(q_{x},q_{j}$), 
	
including ($j_{1}, ..., j_{i-1}$),\end{center}
\end{displayquote}
such an \textit{LCM} would \emph{not} be able to reach the proof object,
\begin{displayquote} \begin{center}
	$\Pi$ $\_$:$\Ga$($\_$). $\Sigma.I(x,$ $\_$) \end{center}
\end{displayquote}
Such would entail that the transition function of \textit{LCM} would execute the computation of a \emph{nondeterministic} machine (no, not “simulate” an \textit{NTM}, but \emph{compute along a nondeterministic path} or \emph{emulate} an \textit{NTM}). 
This much is clearly a contradiction.

Recall Remark 1.9. A \textit{CM} does not merely “simulate” its embedded machines. 
Obviously, if this were the case, it would make little difference whether the \textit{CM} possessed a transition \emph{function} or \emph{relation}, given its corresponding embedded machines (more precisely speaking, it would no longer have embedded machines). 
On the contrary, the embedded machines of a \textit{CM} have computations which accord with the transition function or relation of their “host machine,” \textit{CM}. 
When a \textit{CM} executes the program \textbf{SP}, its transition function/relation “acts as” that of its embedded machine. 
Although it is possible for a \textit{DTM} to simulate an \textit{NTM}, it is not possible for a \textit{DTM} to perform the operation of state transition \emph{relation}, as opposed to \emph{function}.

Note also that bizarre “combined Ceiling Machines” with both deterministic and nondeterministic machines in their extension ($\La$’ = $\La \cup M_{i}, ..., M_{n}$) are indeed permitted. 
In such a case, one would simply need to make the relevant distinctions between the specific means of computation at use at any given time by the Ceiling Machine (transition \emph{function} or transition \emph{relation}). 
At any rate, such oddities do not impair our main result. \begin{displayquote} \end{displayquote}
	\textbf{Remark 3.4.} As one final remark, it may be objected (sloppily!) that \textit{Rice’s theorem} impairs our results. 
This is not the case, however. 
When our \textit{LCM} reads the input corresponding to \textit{CNDS} (the graph of computations of an \textit{HCM}) it does \emph{not} decide on any properties of \emph{CNDS itself}. 
Rather, our \textit{LCM} is given the query as to whether, \emph{through its own programming}, it can reproduce the computations of \textit{CNDS}. 
Our \textit{LCM} simply \emph{searches its memory} for programs corresponding to formulae (proofs) $\varphi_{1}, ..., \varphi_{n}$ of its $\Ta$. 
No “semantic properties” are involved.
\newpage
\epigraph{—Although the specific formal maneuvers of this proof are my own original work, I owe a tremendous debt of conceptual and inspirational gratitude to Christopher Michael Langan. I wish for this proof to be seen as an instance of the metaformal paradigm in action. And I urge readers to study his Cognitive-Theoretic Model of the Universe.}

\epigraph{—I would also like to acknowledge Alexandra Elbakyan’s Sci-Hub. Without this tool of hers, this work would have never made it to its current form. Sci-Hub is an invaluable tool for the “underdog,” and it must not be silenced.}

\epigraph{—Additional thanks are in order for Fionn Curran. Fionn was a tremendous help for the formatting and delivery of this paper.}

\newpage
\nocite{*}
\printbibliography[title={Bibliography}]

\end{document}